\documentclass{article}
\usepackage[a4paper, total={160mm, 9in}]{geometry}
\usepackage{graphicx,amsmath,setspace}
\usepackage[numbers]{natbib}
\usepackage{doi}
\usepackage[T1]{fontenc}
\usepackage[acronym]{glossaries}
\usepackage{hyperref,authblk}

\newacronym{mnte}{MnTe}{manganese telluride}
\newacronym{UHV}{UHV}{ultra-high vaccum}
\newacronym{MBE}{MBE}{molecular beam epitaxy}
\newacronym{ML}{ML}{monolayer}
\newacronym{BL}{BL}{bilayer}
\newacronym{STM}{STM}{scanning tunnelling microscopy}
\newacronym{ARPES}{ARPES}{angle-resolved photoemission spectroscopy}
\newacronym{DFT}{DFT}{density functional theory}
\newacronym{XPS}{XPS}{x-ray photoelectron spectroscopy}
\newacronym{mnse}{MnSe}{manganese selenide}
\newacronym{XMCD}{XMCD}{x-ray magnetic circular dichroism}
\newacronym{XAS}{XAS}{x-ray absorption spectroscopy}
\newacronym{DMI}{DMI}{Dzyaloshinskii-Moriya}
\newacronym{3D}{3D}{three-dimensional}
\newacronym{2D}{2D}{two-dimensional}
\newacronym{nias}{NiAs}{nickel arsenide}
\newacronym{LEED}{LEED}{low-energy electron diffraction}
\newacronym{CVD}{CVD}{chemical vapour deposition}
\newacronym{bite}{Bi$_2$Te$_3$}{bismuth telluride}

\newcommand{\Fig}[2]{Fig. #1#2}
\newcommand{\FigSI}{Fig. S\refstepcounter{SI}\theSI}
\newcounter{SI}
\newcommand{\TableSI}{Table S\refstepcounter{TableSI}\theTableSI}
\newcounter{TableSI}

\begin{document}
\onehalfspacing

\title{Emergent Magnetic Structures at the 2D Limit of the Altermagnet MnTe}


\author[1,2,*]{Marc G. Cuxart}
\author[3,*]{Roberto Robles}
\author[1]{Beatriz Muñiz Cano}
\author[4]{Pierluigi Gargiani}
\author[4]{Clara Rebanal}
\author[4,7,8]{Iolanda Di Bernardo}
\author[4]{Alireza Amiri}
\author[1]{Fabián Calleja}
\author[1,6]{Manuela Garnica}
\author[1]{Miguel A. Valbuena}
\author[1,5,6,7]{Amadeo L. Vázquez de Parga}

\affil[1]{\textit{Instituto Madrileño de Estudios Avanzados, IMDEA Nanociencia, 28049 Madrid, Spain}}
\affil[2]{\textit{Catalan Institute of Nanoscience and Nanotechnology (ICN2), CSIC and BIST, Bellaterra, 08193 Barcelona, Spain}}
\affil[3]{\textit{Centro de Física de Materiales CFM/MPC (CSIC-UPV/EHU), 20018 Donostia-San Sebastián, Spain}}
\affil[4]{\textit{Departamento de Física de la Materia Condensada Universidad Autónoma de Madrid 28049 Madrid, Spain}}
\affil[5]{\textit{ALBA Synchrotron Light Source, 08290 Cerdanyola del Vallès, Spain}}
\affil[6]{\textit{Instituto Nicolás Cabrera, Universidad Autónoma de Madrid, 28049 Madrid, Spain}}
\affil[7]{\textit{Condensed Matter Physics Center (IFIMAC) Universidad Autónoma de Madrid, 28049 Madrid, Spain}}
\affil[8]{\textit{School of Physics and Astronomy - Monash University, Clayton, VIC 3800 Australia}}
\affil[*]{marc.gcuxart@icn2.cat, roberto.robles@csic.es}

\date{}

\maketitle


\begin{abstract}
\acrshort{mnte} has recently emerged as a canonical altermagnet, a newly identified class of magnetism characterized by compensated antiferromagnetic order coexisting with spin-split electronic bands, traditionally considered exclusive to ferromagnets. However, the extent to which altermagnetism persists as altermagnets are thinned to the \gls{2D} limit remains unexplored. Here, we investigate the magnetic behaviour of \gls{2D} \acrshort{mnte}, specifically atomically-thin \glspl{ML} and \glspl{BL} grown on graphene/Ir(111) substrate, by combining experimental \acrlong{STM}, \acrlong{XPS}, \acrlong{XAS} and \acrlong{XMCD} with \acrlong{DFT} calculations. We find that while \gls{ML} and \gls{BL} \acrshort{mnte} adopt atomic structures with symmetries incompatible with altermagnetism, they exhibit intriguing magnetic phases: the \gls{BL} forms a highly-robust layered antiferromagnet with in-plane spin anisotropy, whereas the \gls{ML} exhibits a spin-glass–like behavior below its freezing temperature, a phenomenon not previously observed in an atomically thin material. These findings highlight how reduced dimensionality can promote the emergence of unusual magnetic structures distinct from those of their 3D counterparts, providing new insights into low-dimensional magnetism.
\end{abstract}


\section{Introduction}

\Gls{mnte} has recently attracted significant interest as one of the first materials where altermagnetism has been experimentally confirmed \cite{krempasky2024,zhu2024,fedchenko2024}. Altermagnets are characterized by the coexistence of time-reversal symmetry breaking, leading to spin-split energy bands, and compensated anti-parallel magnetic order \cite{smejkal2022b,smejkal2022}. These were traditionally thought to be exclusive of ferromagnets and antiferromagnets, respectively. Such unique combination grants altermagnets with fundamental and applied interest, as they can host exotic quantum phenomena, including the anomalous Hall effect, spin current and torque generation, and piezomagnetic effects \cite{gonzalezbetancourt2023,bai2023,aoyama2023}.

$\alpha$-\gls{mnte} adopts a hexagonal \gls{nias} structure in its bulk single-crystal form, and typically exhibits collinear antiferromagnetic order with in-plane spin anisotropy, along with high Néel temperature of 310 K. \cite{kunitomi1964,ferrer-roca2000} However, the behaviour of \gls{mnte} at the \gls{2D} limit remains unexplored, where reduced dimensionality and substrate interactions—such as epitaxial strain and domain morphology—can alter its structural, electronic, and magnetic properties. Recent investigations of \gls{mnte} films with tens to hundreds of nanometers thickness have revealed some deviations from the magnetic behaviour observed in bulk crystals, when grown on different substrates. These include decrease of the spin-flop field \cite{kriegner2017a}, the emergence of near-interface ferromagnetic or vortex-like magnetic phases \cite{watanabe2018,amin2024}, and a distinct magnetization behaviour \cite{kluczyk2024}. Yet, altermagnetic manifestations persist even in such \gls{mnte} thin films \cite{gonzalezbetancourt2023,hariki2024,krempasky2024}. This naturally raises the fundamental question of whether altermagnetism persists at the atomically thin limit or, instead, other magnetic structures emerge.

In this work, we investigate the structural, electronic, and magnetic properties of atomically-thin \gls{ML} and \gls{BL} \gls{mnte}, grown on a graphene/Ir(111) substrate via \gls{MBE}, using a combined experimental and computational approach. A comprehensive analysis employing \gls{STM}, \gls{XPS}, \gls{XAS}, \gls{XMCD} and \gls{DFT} calculations reveals that both \glspl{ML} and \glspl{BL} adopt atomic structures distinct from that of bulk single-crystal \gls{mnte}, with symmetries that are incompatible with the persistence of altermagnetism. Notably, \gls{XMCD} measurements indicate that these atomically-thin layers retain antiferromagnetically ordered phases, with \glspl{ML} forming a compensated non-colinear magnet due to magnetic frustration, and \glspl{BL} exhibiting a highly-stable layered in-plane antiferromagnetism.


\section{Morphology and chemical structure of \gls{ML} and \gls{BL} \gls{mnte}}

In our experiments, \gls{mnte} \glspl{ML} and \glspl{BL} were selectively grown by co-deposition of elemental Mn and Te onto a pre-heated graphene/Ir(111) substrate via \gls{MBE} under \gls{UHV} conditions (see Methods section). By varying the deposition time, samples with predominant prevalence of either \glspl{ML} or \glspl{BL} were attained. A representative \gls{ML} \gls{mnte} is shown in the \gls{STM} image of \Fig{1}{a}, visible as a triangle-shaped island and with apparent height of around $3$ Å at low voltage biases. This value is consistent with the thickness reported in the only experimental study of \gls{ML} \gls{mnte} \cite{ding2022}, but slightly thinner than that expected for \gls{ML} MnTe$_2$ \cite{chen2020}. In the uncovered areas surrounding the \gls{mnte} domain, the characteristic moiré pattern that \gls{ML} graphene forms on Ir(111) \cite{busse2011} is visible, thus evidencing the atomically-thin nature of the graphene layer that supports \gls{mnte}. Interestingly, the moiré lattice extends beneath the \gls{mnte} (see \FigSI) and it is still discernible through it, suggesting that structure of \glspl{ML} \gls{mnte} conforms geometrically on graphene. Atomically-resolved \gls{STM} images taken on a flat \glspl{ML} \gls{mnte} domain (\Fig{1}{b,c}) corroborate this hypothesis, as they reveal a bias-dependent contrast inversion induced by the interface interaction between graphene and Ir(111) \cite{voloshina2013}. Rotated by $\sim 19^\circ$ from the moiré lattice, the in-plane atomic structure of the \gls{ML} \gls{mnte} is visible as an hexagonal lattice with a surprisingly large lattice parameter of $4.6 \pm 0.1$ Å, as compared to that of bulk $\alpha$-\gls{mnte} ($4.19$ Å) \cite{efremdsa2005}. Such discrepancy can be attributed to strong epitaxial strain induced by the substrate, but we reject this explanation due to the lack of direct commensurability and huge lattice mismatch between \gls{ML} \gls{mnte} and graphene. Instead, it is consequence of a structural change of $\alpha$-\gls{mnte} when thinned to the \gls{ML} limit, as it is argued below.


\gls{DFT} calculations on free-standing \gls{ML} reveal that the most stable atomically thin \gls{mnte} adopts a planar honeycomb structure of alternating Mn and Te atoms, with a lattice parameter of $4.616$ Å (inset in \Fig{1}{d}), in good agreement with the hexagonal lattice observed experimentally (\Fig{1}{}). The optimized structure of \glspl{ML} \gls{mnte} is connected to that of graphene/Ir(111) through $ \big(\begin{smallmatrix} 6 & 2 \\ -2 & 4 \end{smallmatrix}\big) $ when using the experimentally determined unit cell. In addition, the reported \glspl{ML} \gls{mnte} by Ding \textit{et al.}, which was grown on \gls{bite}, shares the same structure as in the present study \cite{ding2022}. Therefore, we can conclude that the observed structure of \glspl{ML} \gls{mnte} is intrinsically stable, independently from the substrate. The concordance between experiment and calculations is further supported by the close agreement between measured and simulated \gls{STM} images of the substrate-supported \gls{ML} \gls{mnte}, including the vanishing contrast originated from the underlying graphene/Ir(111) moiré pattern at low bias voltages (\Fig{1}{b,c}). Based on this structural analysis, the degree of interaction between \gls{ML} \gls{mnte} and substrate is expected to be negligible, leading to a largely unaffected properties of \gls{ML} \gls{mnte}. Such minimal interface interplay is corroborated experimentally by the easy move of \gls{mnte} islands with the \gls{STM} tip (\FigSI), and theoretically by the tiny energy differences between different \gls{mnte} - substrate registry configurations (\FigSI). Indeed, minute changes in the magnetic properties are calculated for the isolated and substrate-supported \gls{ML}, as it will be discussed in section \ref{magnetism}.


Chemical \gls{XPS} characterization of samples with predominant prevalence of \gls{ML} \gls{mnte}, as shown in the \gls{STM} image in \Fig{2}{a}, supports the proposed Mn$_1$Te$_1$ honeycomb structure, in which Mn should become doubly charged (Mn$^{+2}$) upon coordination with Te. First, the Mn $2p$ core-level signal exhibits the characteristic line-shape of Mn$^{+2}$, consisting of a complex multiplet splitting and a shake-up resonance, which can be modelled by fitting the series of sub-peaks labelled as 'Mn 2+' according to reference \cite{biesinger2011}, as shown in \Fig{2}{b}. Second, Mn:Te relative concentration extracted form the \gls{XPS} intensities of Mn $2p$ and Te $3d$ core-level peaks accounts for $44:56$, in fairly good agreement to the expected $1:1$ stoichiometry. For the sake of completeness, \gls{XPS} C 1s and Ir 4f core levels are provided in the supplementary information (\FigSI).

Upon increase of co-deposition time of Mn and Te, domains with a thickness twice that of the \gls{ML} ($\sim 7$ Å, \Fig{2}{d}) are identified as \gls{mnte} \glspl{BL}. These exhibit a lattice parameter ($4.6 \pm 0.1$ Å) equal to that of the \gls{ML} within the resolution limit, and also mimic the underlying moiré lattice of graphene/Ir(111). By applying the same peak analysis procedure as for the \gls{ML} case, the \gls{XPS} Mn 2p spectra of \gls{BL} \gls{mnte} (\Fig{2}{e}) shows that Mn remains in a +2 charge state. The extracted Mn:Te relative concentration in this case ($44:56$) is close to that expected for a 1:1 stoichiometric \gls{mnte}, as argued	in the following paragraph. Note that the slight Te excess in nominally \gls{BL} \gls{mnte} samples may be attributed to the coexistence of \glspl{BL} with Te adatoms intercalated between graphene and Ir(111), as observed by \gls{STM} in \FigSI~ \cite{cano2023}.

\gls{DFT} calculations reveal that the \gls{BL} structure consists of two gently buckled honeycomb \gls{mnte} \glspl{ML} stacked in an AB configuration, where Mn atoms are bonded to neighbouring Te atoms forming trigonal prismatic geometry (\Fig{3}{a}), analogous to that of \gls{mnse} \cite{qayyum2024}. Similar to the case of the \gls{ML}, this \gls{BL} structure is found to be the most stable solution in both graphene/Ir(111)-supported and free-standing calculations, thus highlighting the minimal influence of the substrate into \gls{BL} \gls{mnte}. This is further supported by the preserved free-standing structural and magnetic properties of the \gls{BL}, as calculated by \gls{DFT} (discussed below), and the weak interaction evidenced by the easy tip-induced displacement of \gls{BL} domains. Finally, the calculated in-plane lattice parameter ($4.434$ Å) is smaller than that of ML, but still agrees with the measured value for the BL within $5 \%$. Similarly, the calculated C--(upper) Te distance, which ranges from $7.234$ Å to $7.332$ Å depending on the moiré site \Fig{3}{b}), is consistent with the apparent thickness observed by \gls{STM}.

%

\section{Magnetism of \gls{ML} and \gls{BL} \gls{mnte}}\label{magnetism}
We employed synchrotron-based x-ray absorption techniques, namely \gls{XAS} and \gls{XMCD} to investigate the magnetic structure of \gls{ML} and \gls{BL} \gls{mnte} with element and orbital sensitivity. These measurements were conducted in a correlative approach involving \gls{STM}, \gls{XPS} and \gls{XMCD}, with samples transferred between measurement setups via a \gls{UHV} suitcase (see section \ref{methods}). \Fig{4}{a} depicts \gls{XAS} spectra measured at the Mn $L_{2,3}$ absorption edge, on the sample with predominant concentration of \gls{ML} \gls{mnte}, under an applied external magnetic field of $6$ T parallel to the incident photon beam and at $\sim 3.5$ K (as shown in the small schematics in \Fig{4}{a,b}). The complex edge shape of Mn L$_3$, corresponding to the dipole-allowed transitions from the spin-orbit split Mn 2p$_{3/2}$ to the 3d unoccupied states, exhibits multiplet features around a main peak at $639.6$ eV, characteristic of Mn in a +2 state \cite{gilbert2003}. This is in agreement with the charge state determined by \gls{XPS} in \Fig{2}{b}. The emergence of the small $L_3$ shoulder at $638.5$ eV and the triple-shaped $L_2$ edge substructure, neither of them present in bulk \gls{mnte} \cite{awana2022,hariki2024}, denotes that the Mn 2+ cations are subject to a moderately more intense crystal field acting on its 3d shell \cite{degroot1990}. This is consistent with the reduced Mn--Te distance calculated for the \gls{ML} structure ($2.691$ Å) compared to that for the bulk structure ($2.912$ Å).

\gls{XAS} spectra were measured at normal ($\theta = 0^\circ$, in \Fig{4}{a}) and grazing incidence ($\theta = 70^\circ$, in \FigSI) using right ($C_+$) and left ($C_-$) circularly polarized light. This configuration enables access to the difference in spin population projected along the out-of-plane and in-plane directions, respectively. These are represented by the \gls{XMCD} signals in \Fig{4}{b}, expressed in percentage of the average \gls{XAS} (black line in \Fig{4}{a}). The \gls{XMCD} minimum at $639.6$ eV indicates that the Mn magnetic moments are primarily carried by the 3d shell, while the very similar angle-dependent magnetization (approximately $-12 \%$) evidence the lack of magnetic anisotropy in \gls{ML} \gls{mnte}. This result contrasts with the well-defined in-plane anisotropy of the antiferromagnetic order present in bulk \gls{mnte} \cite{kunitomi1964}. Applying magneto-optical sum rules\cite{thole1992,carra1993} yields small but sizeable induced spin and orbital moments of $\mathbf{m_s} = (0.121 \pm 0.003) \ \mu_B$/Mn atom and $\mathbf{m_l} = (0.010 \pm 0.0008) \ \mu_B$/Mn atom, respectively (both calculated in the exemplary spectra measured in the out-of-plane configuration). While the absolute values are of limited relevance given that they have been measured far from magnetic saturation point (see the magnetization curve in \Fig{4}{c}), they confirm that the tiny induced magnetization is primarily of spin origin, with minimal orbital contribution, as expected. This is consistent with an effect of field-induced magnetic canting \cite{mazin2024}, as it is rationalized in the discussion section.

The magnetization curves presented in \Fig{4}{c} further support the dominant isotropic magnetic behaviour of \gls{ML} \gls{mnte}, as indicated by the quite similar line-shape and magnetization values reached at $\pm 6$ T in the in-plane and out-of-plane measurements. These curves were obtained after measuring the \gls{XMCD} signal at sequentially decreasing magnetic fields (from $6$ T to $-6$ T) at $\sim 3.5$ K, under both normal and grazing configurations. Additionally, the curves reveal that the magnetization increases smoothly with the applied magnetic field, without reaching saturation, and that no remnant magnetization is detected above the resolution limit. In this scenario, a paramagnetic response can be ruled out, as the expected Mn spin state configuration of 5/2 would yield a larger magnetization under the field and temperature conditions, see the discussion section. Instead, the observed behaviour is consistent with a canted compensated magnetic system, where the induced moment arises from a field-dependent spin canting rather than intrinsic ferromagnetic ordering \cite{mazin2024}.

The same approach followed for magnetic characterization of \gls{ML} \gls{mnte} yielded significantly different results when applied to the \gls{BL} \gls{mnte}. The Mn L$_{2,3}$ edge line shape, as measured by \gls{XAS} at normal (\Fig{5}{a}) and grazing (Fig. S6) incidence, is still compatible with Mn$^{+2}$ but more closely resembles that of bulk \gls{mnte} than that of the \gls{ML} \cite{awana2022,hariki2024}. It exhibits an L$_2$ peak with less pronounced low-energy shoulder and maximum positioned at slightly lower energy ($639.2$ eV), which is indicative of different contribution of the crystal field resulting from the inequivalent Mn coordination environment in the \gls{BL} and \gls{ML} structures \cite{degroot1990,hariki2024}. On the other hand, the \gls{XAS} spectra measured with $C_+$ and $C_-$ photons appear with nearly identical shape (\Fig{5}{a}) resulting in a barely detectable \gls{XMCD} signal of more than one order of magnitude smaller than that of \gls{ML}, even at $6$ T (\Fig{5}{b}). An equivalent behaviour is observed for the spectra taken at grazing measurement geometry, in consistency with a rather isotropic system. Such a tiny dichroic signal may result from residual \gls{ML} domains, as observed by \gls{STM} in \Fig{2}{d}, or from regions that may contain intercalated Mn adatoms between graphene and Ir(111). 

\section{Discussion}

To interpret the experimental findings from \gls{XMCD}, we performed \gls{DFT} calculations considering different magnetic configurations for graphene/Ir(111)-supported \gls{ML} and \gls{BL} \gls{mnte} (see \FigSI~ and \FigSI). To validate our \gls{DFT} model, it was first applied to the \gls{nias}-type atomic structure of 3D-bulk \gls{mnte}, where it correctly predicted the expected interlayer antiferromagnetism with in-plane anisotropy and magnetic moment per Mn atom (\FigSI) \cite{kriegner2017a}.

For the case of \gls{ML} \gls{mnte}, the most stable magnetic configuration yields a magnetic moment of $4.75 \ \mu_B$ per individual Mn atom, as expected for a 2+ high-spin configuration of the Mn cations in the \gls{ML}. However, this solution depicts a non-collinear configuration (\Fig{4}{c}), which emerges due to the intrinsic frustration of antiferromagnetic coupling between nearest Mn atoms in their hexagonal lattice of the \gls{ML} \gls{mnte}. In this non-collinear configuration, the global Mn magnetization becomes fully compensated ($< 0.01 \ \mu_B$/Mn atom), in agreement with the absence of net magnetization observed in the \gls{XMCD} measurements of the Mn L$_{2,3}$ at $0$ T magnetic field (\Fig{4}{b}).
Different non-collinear solutions coexist within a small energy range of $\sim 1$ meV (Fig. S7). The existence of almost degenerate non-equivalent non-collinear solutions,  together with the absence of magnetic saturation even at $\pm 6$ T (\Fig{4}{c}) are characteristics compatible with a spin glass below the freezing temperature \cite{binder1986}. If confirmed, to the best of the authors knowledge, this would make \gls{ML} \gls{mnte} the first material to exhibit spin-glass behaviour at the atomically-thin limit \cite{pal2024,fritsch2017}. Further experiments are needed to fully confirm the spin-glass nature of \gls{ML} \gls{mnte}.

It is worth noting that the weak increase of magnetization observed upon an applied magnetic field may be attributed to field-induced spin canting. This has been recently observed in bulk \gls{mnte} and interpreted as \gls{DMI}-driven weak ferromagnetism \cite{mazin2024}.

Moving to the \gls{BL} \gls{mnte}, the most stable free-standing solution features antiferromagnetic interlayer Mn-Mn coupling, with Mn atoms within each layer coupling ferromagnetically (\Fig{5}{c}), and exhibits a strong in-plane magnetocrystalline anisotropy of $-0.77$ meV/Mn atom. Notably, the computed energy difference between the most stable antiferromagnetic and ferromagnetic solutions is extraordinarily large, approximately $3.85$ times that of bulk \gls{mnte}, making \gls{BL} \gls{mnte} an exceptionally stiff antiferromagnet. This substantial energy difference is attributed to its distinct atomic structure, which imposes smaller Mn--Mn interlayer distances compared to bulk \gls{mnte}. Similar energy values were reported for \gls{mnse} in the computational study by Qayyum \textit{et al}. \cite{qayyum2024}, which shares the same unit cell as \gls{mnte}.

The incorporation of the graphene/Ir(111) substrate into the calculation does not significantly alter the behaviour of \gls{BL} \gls{mnte}. The interlayer in-plane antiferromagnetic ordering remains the most stable configuration \Fig{5}{c}, with a slightly reduced exchange coupling and magnetic anisotropy energies compared to the free-standing \gls{BL}. Yet, these values remain several times larger than those of bulk \gls{mnte} (see \TableSI). Such robust antiferromagnetism explains the absence of magnetization up to $6$ T, as observed experimentally by \gls{XMCD}.

Finally, we assess whether the altermagnetic behavior of \gls{mnte} persists in its mono- and bi- atomically thin limit. \Gls{3D} altermagnets, such as bulk \gls{mnte}, exhibit a compensated magnetic order where opposite-spin sublattices are connected by crystal-rotation symmetries rather than inversions or translations \cite{osumi2024,smejkal2022}. Since the moment compensation arises from spin-symmetry principles, the classification of \gls{ML} and \gls{BL} \gls{mnte} within this magnetic phase can be determined by analysing their crystal and spin lattices \cite{smolyanyuk2024}. In \gls{ML} \gls{mnte}, the antiferromagnetic order highly deviates from collinearity due to magnetic frustration in the hexagonal Mn lattice. This differs from the collinear spin alignment required for altermagnets, thereby classifying \gls{ML} \gls{mnte} as an unusual non-collinear antiferromagnet. On the other hand, the two spin sub-lattices residing in opposite Mn planes in \gls{BL} \gls{mnte} can be connected by a single inversion transformation, an operation not compatible with the symmetry constraints of altermagnetism. Moreover, it has been shown that in this structure, time-reversal symmetry is globally conserved \cite{qayyum2024}, thereby satisfying Kramer degeneracy principle and suppressing the emergence of altermagnetic band splitting. Hence, \gls{BL} \gls{mnte} can be classified as an exceptionally robust layered antiferromagnet rather than an altermagnet. In conclusion, our results demonstrate how symmetry breaking and magnetic frustration in the two-dimensional limit of the canonical altermagnet \gls{mnte} give rise to unexpected spin textures, establishing atomically-thin \gls{mnte} films as model systems for both fundamental studies of low-dimensional magnetism and the development of antiferromagnetic spintronic devices.




\section{Methods} \label{methods}
\textbf{Samples preparation}. The two types of samples studied in this work, those containing majority of \gls{mnte} \glspl{ML} and \glspl{BL} were grown by \gls{MBE} on Ir(111)-supported graphene in a \gls{UHV} chamber as it follows. First, an atomically clean and flat Ir(111) surface was obtained after repeated cycles of Ar$^+$ sputtering ($3 \times 10^{-6}$ Torr, $1$ keV) and annealing at $1400$ K. Second, a \gls{ML} graphene was grown on Ir(111) by \gls{CVD} of $33$ L of ethylene on pre-heated Ir(111) at $1400$ K. Then, \gls{mnte} layers were grown on the graphene/Ir(111) substrate by co-evaporation of elemental Te (MaTecK, 99.9999 \% purity) and Mn (MaTecK, 99.8 \% purity), from a Knudsen (T = $600$ K) and an e-beam heated cell ($1$ kV, $\sim 25$ nA), respectively. Selectivity over the dominant prevalence of \glspl{ML} and \glspl{BL} was attained by controlling the deposition time, $< 30$ and $> 60$ minutes, respectively. Samples were not exposed to the atmosphere during transfer from the \gls{MBE} growth chamber to the \gls{STM} chamber, as they were \emph{in-situ} connected by \gls{UHV}. Neither in the transfer to the \gls{XPS} and \gls{XAS}/\gls{XMCD} chambers, which was conducted using a home-built \gls{UHV} suitcase kept at a base pressure of $5 \times 10^{-9}$ mbar.

\textbf{\gls{STM} measurements} were conducted in an Omicron variable-temperature \gls{STM} operating at room temperature and \gls{UHV} conditions. The \gls{STM} head was operated with a Nanonis electronics. Bias voltages given in the manuscript refer to the sample voltage. $X, Y$ piezoelectric scanners were calibrated using the lattice parameter of graphene/Ir(111) moiré structure ($24.61$ Å)\cite{meng2012}, while $Z$ using the monoatomic step of Ir(111) ($2.22$ Å).

\textbf{\gls{XPS} measurements} were performed with a Phoibos 150 analyzer and XR-50 x-ray source (SPECS GmbH, Berlin, Germany) operating at room temperature and base pressure $<10^{-10}$ mbar, $10$ eV pass energy (unless stated otherwise), with Al K$_\alpha$ ($h \nu$ = $1486.6$ eV) as photon-source and at normal emission geometry.

\textbf{\gls{XAS} and \gls{XMCD} measurements} were carried out at BOREAS  beamline of the ALBA Synchrotron facility \cite{barla2016}. Spectra were taken in  total-electron-yield mode with right ($C_+$) and left ($C_-$) circularly polarized photons at normal ($0^{\circ}$) and grazing ($70^{\circ}$) incidence, in the presence of a magnetic field up to $\pm 6$ T aligned parallel to the incident beam and at a sample temperature and base pressure of $\sim 3.5$ K (cold finger at $1.5$ K) and $1 \times 10^{-10}$ mbar, respectively.

\textbf{\gls{DFT} calculations} were performed using the Vienna \textit{Ab initio} Simulation Package (VASP) \cite{kresse1996}. For exchange and correlation, the PBE form of the GGA functional was used \cite{perdew1996}. van der Waals forces in this functional were included by applying the Tkatchenko/Scheffler method \cite{tkatchenko2009}. Core electrons were treated within the PAW method \cite{kresse1999}, while wave functions were expanded using a plane-wave basis set with an energy cutoff of $500$ eV. The GGA+U method \cite{dudarev1998} was applied to treat the Mn $3d$ electrons, with $U_{\text{eff}} = U - J = 3$ eV \cite{kriegner2017a}. Calculations to determine non-collinear configurations and magnetic anisotropy energies were carried out including SOC, as implemented in VASP \cite{steiner2016}. For the simulations of free-standing \gls{ML} and \gls{BL} \gls{mnte}, a $2 \times 1$ unit cell was used to consider antiferromagnetic configurations, with a $6 \times 12 \times 1$ $\Gamma$-centered \textit{k}-grid. The graphene/Ir(111) surface was modelled using a slab with three Ir layers. To account for the moiré pattern, a $10 \times 10$ graphene/$9 \times 9$ Ir supercell was used, following previous studies. \cite{jarvi2022} All atoms except those in the bottom Ir layer were relaxed until the forces were smaller than $0.01$ eV/Å.

\gls{STM} images were simulated using the Tersoff--Hamann method~\cite{tersoff1985}, as implemented in the \gls{STM}pw code \cite{lorente2019}. Due to the size of the unit cell, the $\Gamma$ point was sufficient for structural relaxations, while a $3 \times 3 \times 1$ $\Gamma$-centered \textit{k}-grid was used for the \gls{STM} simulations. Atomic magnetic moments were determined via Bader analysis \cite{tang2009a}. Ball-and-stick models were visualized using the VESTA and VMD programs \cite{momma2011, humphrey1996}.

\section{Acknowledgements}
This work was supported by the Spanish Ministry of Science and Innovation and the European Union, Grant Nos. PID2021-123776NB-C21 (CONPHASETM), PID2021-128011NB-I00. Also by Ministerio de Ciencia, Innovación y Universidades (MICIU/AEI/10.13039/501100011033) through grant, PID2021-128011NB-I00. Comunidad de Madrid through grant “Materiales Disruptivos Bidimensionales (2D)” MAD2D-CM-UAM. Financial support from MAD2D-CM projects — MRR MATERIALES AVANZADOS-IMDEA-NC and MRR MATERIALES AVANZADOS-UAM — is also acknowledged.
IMDEA Nanociencia and IFIMAC acknowledge financial support from the Spanish Ministry of Science and Innovation through the "Severo Ochoa" Programme for Centres of Excellence in R\& D (Grant CEX2020-001039-S) and "María de Maeztu" (Grant CEX2018-000805-M) Programme for Units of Excellence in R\& D, respectively.
M.G.C. acknowledges funding from the European Union’s Horizon 2020 research and innovation programme under the Marie Skłodowska-Curie grant agreement no. 101034431 (IDEAL programme) and financial support from MCIN with funding from European Union NextGenerationEU (PRTR-C17.I1) and Generalitat de Catalunya.
M.G. acknowledges financial support through the “Ramón y Cajal” Fellowship program (RYC2020-029317-I) and the “Ayudas para Incentivar la Consolidación Investigadora” programme (CNS2022-135175).
Financial support through the "Soluciones sostenibles del NanoMAGnetismo para TIC" (Mag4TIC-CM) is acknowledged.
I.D.B. acknowledges funding from the European Union’s Horizon Europe research and innovation programme under the Marie Skłodowska-Curie grant agreement no.101063547 (STORM) and support  from the "Ramón y Cajal" program, grant no. YC2022-035562-I.
We acknowledge beam time on BOREAS (BL29) beam line at ALBA Synchrotron under proposal 2023027388, and beam line scientist Dr. M. Valvidares for his supportive commitment to the project. 

\section{Author Contributions}
M.G.C. and A.L.V. conceived the study. R.R. performed the \gls{DFT} calculations and analysed the computational data. M.G.C., B.M.C and M.A.V. performed the \gls{XPS} measurements. M.G.C., A.A. and M.G performed the \gls{STM} measurements. M.G.C., C.R., I.D.B. and P.G. performed the \gls{XMCD} measurements. M.G.C. analysed the experimental data and wrote the manuscript with feedback from all other authors.

\section{References}
\bibliographystyle{nature}
\bibliography{Main_text_arXiv}

\newpage
\begin{figure}
	\includegraphics{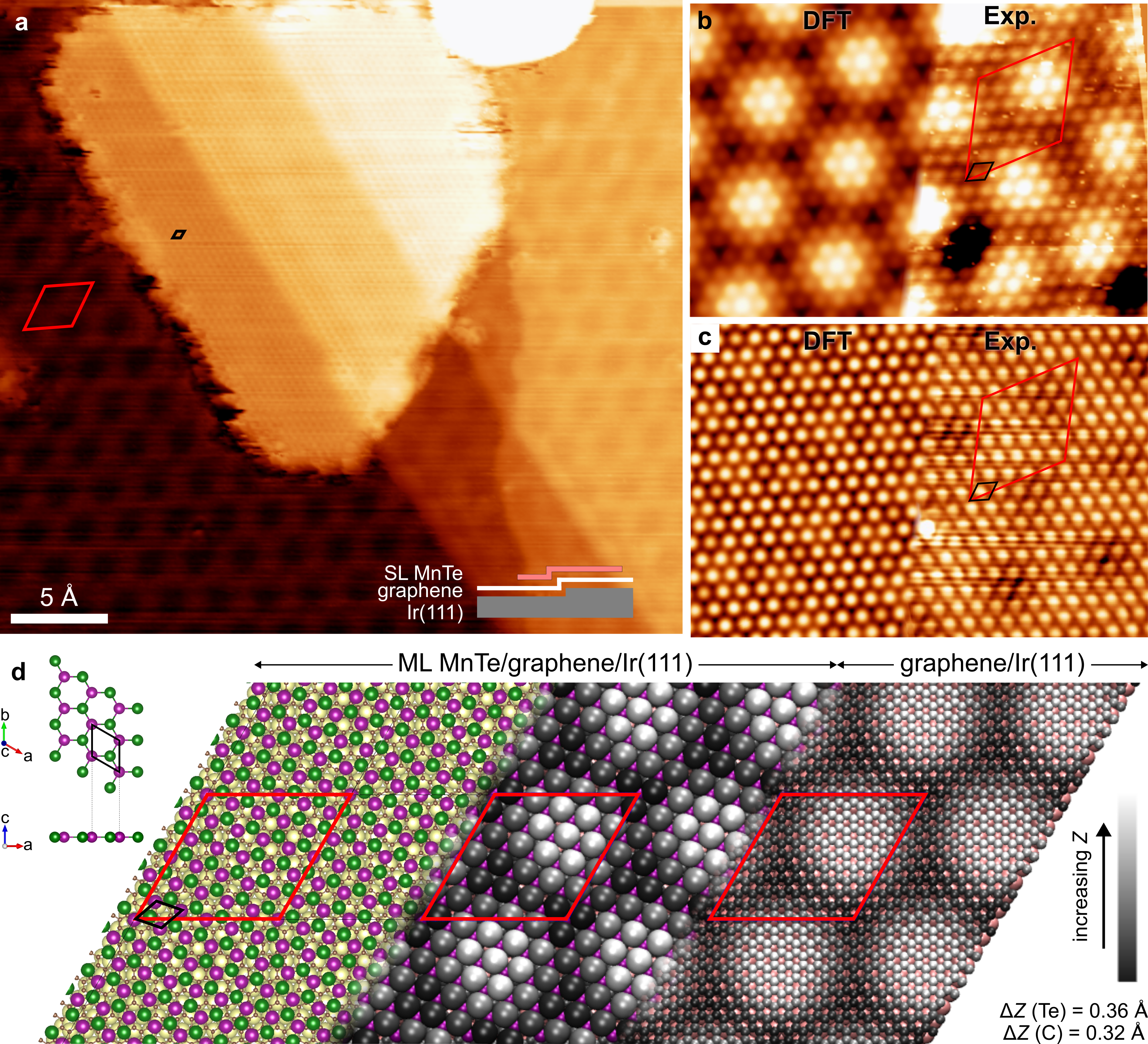}
	\caption{\textbf{Morphology and structure of ML MnTe on graphene/Ir(111).}
	\textbf{a}, STM image of a single-crystalline domain sitting across three atomically flat terraces of the graphene/Ir(111)substrate (scanning parameters $2.0$ V, $0.4$ nA).
	Experimental and simulated atomically-resolved STM images (left and right side of the images, respectively), taken on the same region at \textbf{b} $2.0$ V ($0.4$ nA) and \textbf{c} $0.5$ V ($0.3$ nA).
	\textbf{d}, Ball-and-stick models representing DFT-optimized structures of ML MnTe on graphene/Ir(111) (left and central parts) and graphene/Ir(111) (right part). In the central and right structures, Te and C atoms, respectively, have been color-coded according to their adsorption height ($Z$). Top and side view of the optimized free-standing ML MnTe structure in the top-left inset. Black and red rhomboids represent the unit cell of the atomic lattice of ML MnTe and the moiré lattice of graphene/Ir(111), respectively. Brown and yellow balls represent C and Ir atoms, respectively.
	}
\end{figure}

\begin{figure}
	\centering
	\includegraphics{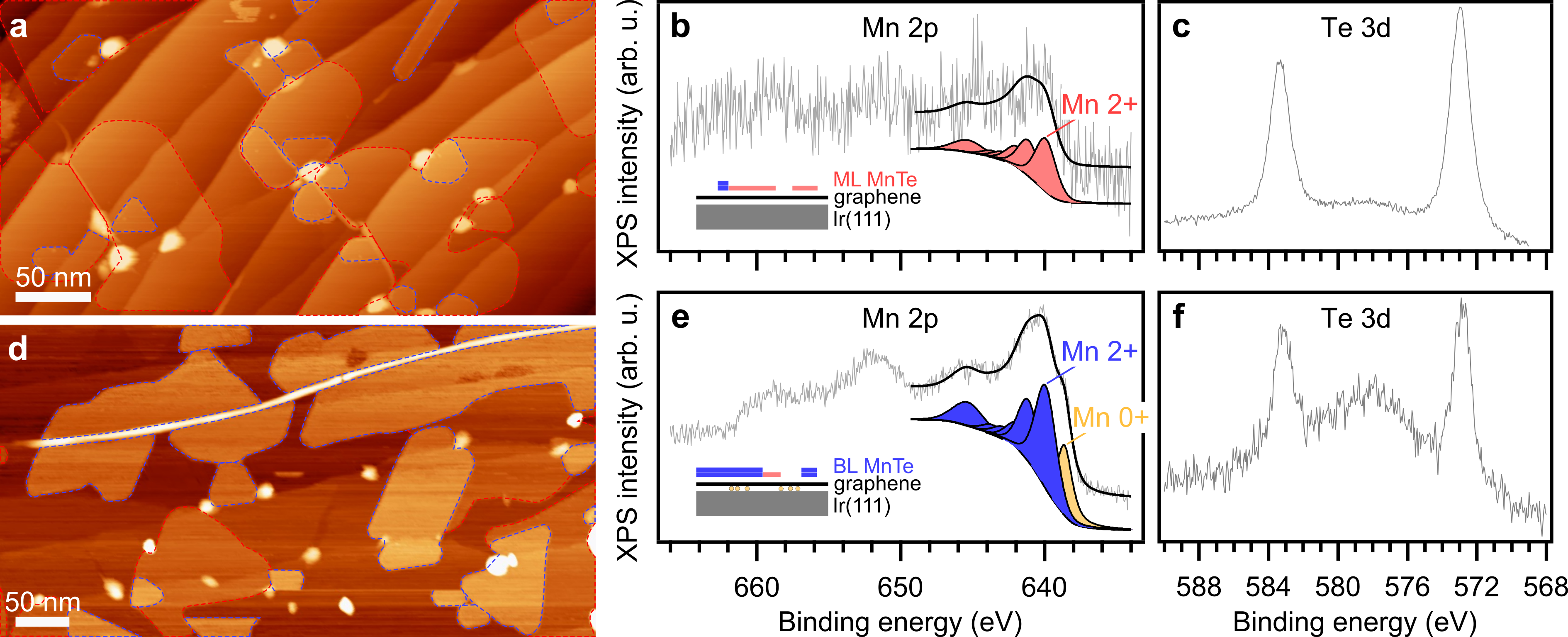}
	\caption{\textbf{Morphology and chemical structure of ML and BL MnTe on graphene/Ir(111).}
	Representative STM images of samples with major prevalence of \textbf{a} MnTe MLs and \textbf{b} BLs (scanning parameters $1.0$ V, $0.3$ nA and $0.5$ V, $0.3$ nA, respectively). ML and BL domains are highlighted by dashed red and blue lines, respectively.
	XPS Mn 2p and Te 3d core levels measured on \textbf{b-c} ML and \textbf{e-f} BL samples. Experimental raw data are shown as thin grey lines, fits of the indicated components as coloured areas, and the sum of the fits as solid black lines. Raw data and sum of the fits are vertically offset for a better visualization. Peak fitting has been converged after fixing the relative intensity and energy position between sub-peak Mn +2 components (in red in \textbf{b} and blue in \textbf{e}) according to reference \cite{biesinger2011}. Only the relative intensities between Mn +2, Mn +0 component (and with the high energy shake-up feature at $\sim 645$ eV) and Shirley background were left free.
	}
\end{figure}

\begin{figure}
	\centering
	\includegraphics{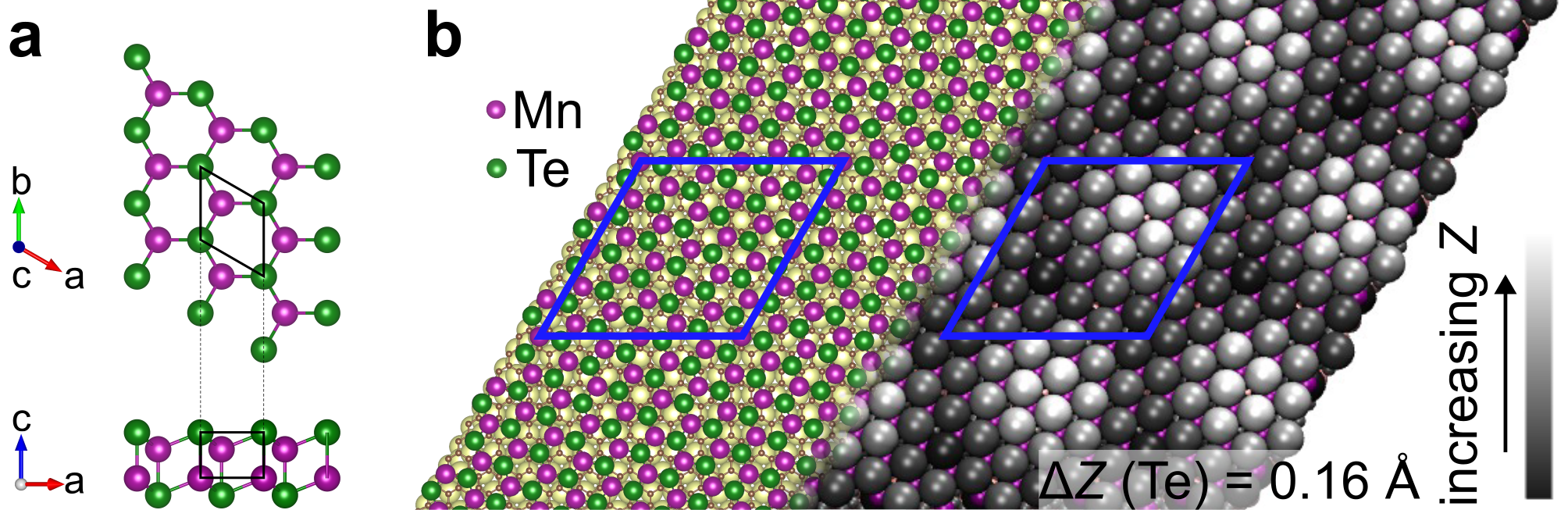}
	\caption{\textbf{a}, Top and side view of the DFT-optimized free-standing BL MnTestructure, represented as a ball-and-stick model.
	\textbf{b}, Ball-and-stick models representing DFT-optimized structures of BL MnTe on graphene/Ir(111). In the right side of the panel, Te atoms have been color-coded according to their adsorption height ($Z$). Black and red rhomboids represent the unit cell of the atomic lattice of BL MnTe and the moiré lattice of graphene/Ir(111), respectively. Brown and yellow balls represent C and Ir atoms, respectively.
	}
\end{figure}

\begin{figure}
	\centering
	\includegraphics{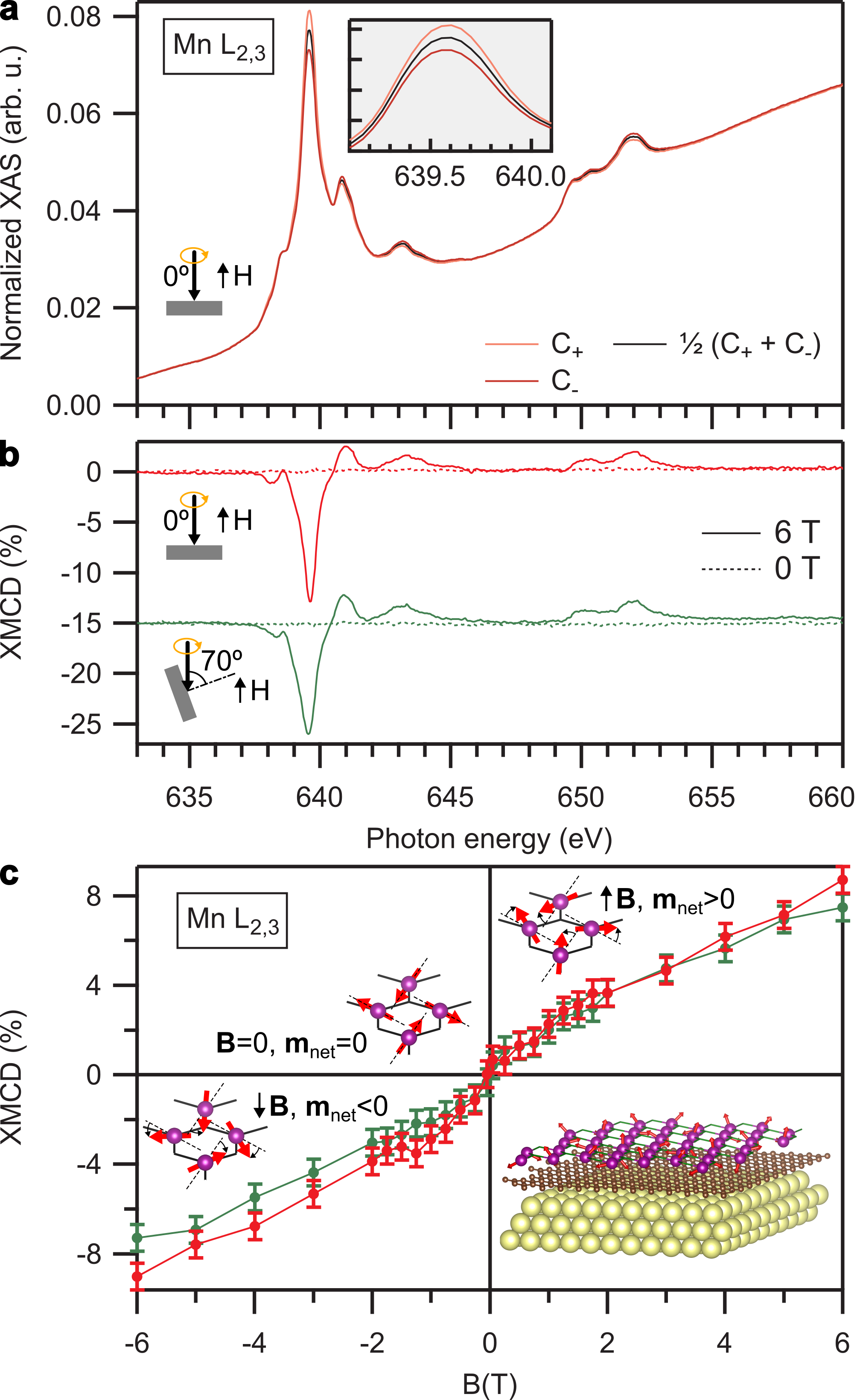}
	\caption{\textbf{Magnetic properties of ML MnTe on graphene/Ir(111).}
	\textbf{a}, Mn L$_{2,3}$ edge measured by XAS on a sample with a major prevalence of MLs, using $C_+$ and $C_-$ polarized light incident at normal angle to the sample surface plane, under a $+6$ T magnetic field. Magnified view of the absorption maximum of the L$_3$ edge.
	\textbf{b}, XMCD spectra obtained by subtracting $C_-$ from $C_+$ XAS, expressed as a percentage of the $C_+$ and $C_-$ average, at normal (red) and grazing (green) incidence, measured at $+6$ T and after removing the magnetic field ($0$ T). The green curve has been vertically offset by $-15$ \%.
	\textbf{c}, Mn magnetization curve measured on the same sample by recording the XMCD signal at decreasing magnetic fields, at normal (red) and grazing (green) incidence. Inset: Ball-and-stick representation of the most stable magnetic configuration featuring non-collinear antiferromagnetic order. Te atoms of MnTe are not represented for clarity of visualization. Red arrows depict the atomic magnetic moments, and purple, brown and yellow balls represent Mn, C and Ir atoms, respectively. All measurements were performed at $\sim 3.5$ K.
	}
\end{figure}

\begin{figure}
	\centering
	\includegraphics{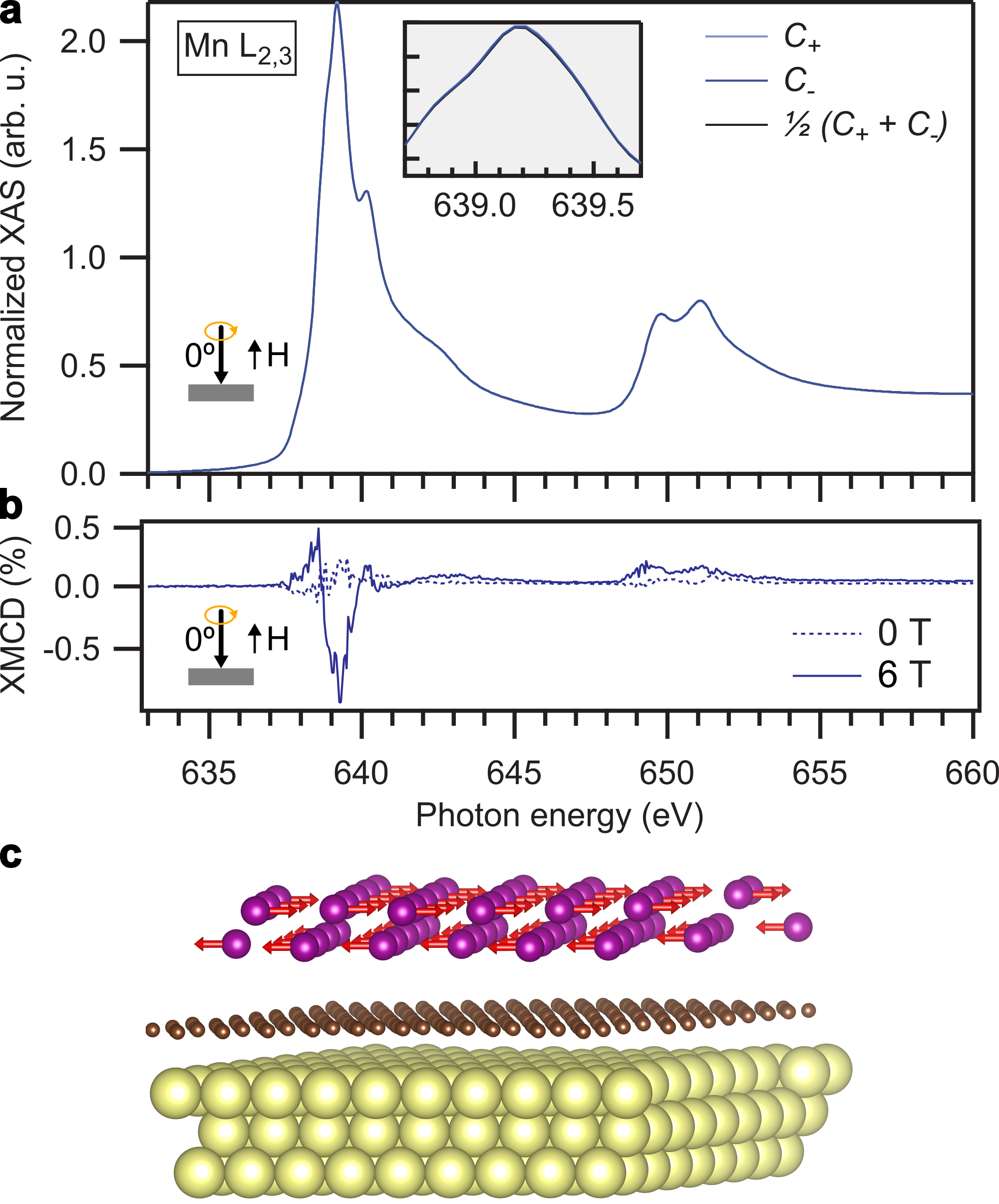}
	\caption{\textbf{Magnetic properties of BL MnTe on graphene/Ir(111)}.
	\textbf{a}, Mn L$_{2,3}$ edge measured by XAS on a sample with a major prevalence of BLs, using $C_+$ and $C_-$ polarized light incident normally to the sample surface plane, under a $+6$ T magnetic field. Magnified view of the absorption maximum of the L$_3$ edge.
	\textbf{b}, XMCD spectra obtained by subtracting $C_-$ from $C_+$ XAS, expressed as a percentage of the $C_+$ and $C_-$ average, measured at normal incidence, at $+6$ T and after removing the magnetic field ($0$ T).
	\textbf{d}, Ball-and-stick representation of the most stable magnetic configuration featuring interlayer ferromagnetic and interlalyer antiferromagnetic order. Te atoms of MnTe are not represented by the sake of visualization. The represented in-plane anisotropy has been assumed to be the same as in the free-standing BL. Red arrows depict the atomic magnetic moments, and purple, green, brown and yellow balls represent Mn, Te, C and Ir atoms, respectively.
	}
\end{figure}

\end{document}